\documentclass[10pt,letter,final]{article}

\usepackage[backref,
            natbib
            ,style = numeric-comp
            ,maxnames = 2
            ,backend = bibtex
            ,sorting=none
            ]{biblatex}
  
\usepackage[T1]{fontenc} 
\usepackage{palatino,eulervm}
\usepackage[english]{babel} 
\usepackage{csquotes}
\usepackage{amsmath,amssymb,amsfonts,amsthm} 
\usepackage[table,dvipsnames*,svgnames]{xcolor}
\usepackage{fixltx2e}
\usepackage{showkeys}
\usepackage{bm} 

\usepackage[top=2.5cm, bottom=2.5cm, outer=3cm, inner=3cm, heightrounded, marginparwidth=2cm, marginparsep=0.5cm]{geometry} 
\usepackage[active]{srcltx}
\usepackage{extarrows}

\usepackage{braket}
\usepackage{mathrsfs}
\usepackage{marginnote}
\usepackage[utf8]{inputenc}

\makeatletter

\let\@authors\@empty
\let\@email\@empty
\let\@affiliationone\@empty
\let\@affiliationtow\@empty
\let\@pdfsubject\@empty
\let\@keywords\@empty
\let\@preprint\@empty

\providecommand{\pdfsubject}[1]{\gdef\@pdfsubject{#1}}
\providecommand{\keywords}[1]{\gdef\@keywords{#1}}
\renewcommand{\author}[1]{\ifx\@authors\@empty\toks@\expandafter{#1}\else\toks@\expandafter{\@authors, #1}\fi\edef\@authors{\the\toks@}}
\providecommand{\email}[1]{\ifx\@email\@empty\toks@\expandafter{#1}\else\toks@\expandafter{\@email, #1}\fi\edef\@email{\the\toks@}}
\providecommand{\affiliationone}[1]{\gdef\@affiliationone{#1}}
\providecommand{\affiliationtwo}[1]{\gdef\@affiliationtwo{#1}}
\providecommand{\preprint}[1]{\gdef\@preprint{#1}}

\makeatother

\usepackage[hyperindex,breaklinks]{hyperref}

\hypersetup{
    unicode=false,          
    pdftoolbar=true,        
    pdfmenubar=true,        
    pdffitwindow=false,     
    pdfstartview={FitH},    
    pdfproducer={SGA, BUWS}, 
    pdftitle={Soft Black Hole Absorption Rates as Ward Identities},
    pdfsubject={Absorption rates of black holes and large gauge transformations},
    pdfkeywords={Black Holes, Absorption, LGT},
    pdfnewwindow=true,      
    colorlinks=true,       
    linkcolor=black,          
    citecolor=DarkGreen,        
    filecolor=magenta,      
    urlcolor=cyan!30!black!70         
}

\usepackage{graphicx}

\allowdisplaybreaks[3]
\newcommand{\del}{\partial}

\newcommand{\bal}{\begin{align}}
\newcommand{\eal}{\end{align}}
\newcommand{\scri}{\mathscr{I}}

\DeclareMathOperator{\tr}{tr}

\renewcommand{\[}{\begin{equation}}
\renewcommand{\]}{\end{equation}}

\newcommand{\drm}{\mathrm{d}}
\newcommand{\pd}{\partial}
\newcommand{\wkeq}{\overset{w}{=}}

\newcommand{\nln}{\notag\\}

\bibliography{bh-low-abs.bib}

\begin{document}

\title{
Soft Black Hole Absorption Rates as Conservation Laws}

\author{Steven G.\ Avery\textsuperscript{a}}
\email{\href{mailto:savery@msu.edu}{savery@msu.edu}}

\affiliationone{%
\textsuperscript{a}Brown University\\Department of Physics\\182 Hope St, Providence, RI 02912\\ 
\vspace{3mm}
\textsuperscript{a}Michigan State University\\Department of Physics and Astronomy\\ East Lansing, MI 48824
}

\author{Burkhard U.\ W.\ Schwab\textsuperscript{b}}
\email{\href{mailto:schwab@cmsa.fas.harvard.edu}{schwab@cmsa.fas.harvard.edu}}

\affiliationtwo{%
\textsuperscript{b}Harvard University\\Center for Mathematical Science and Applications\\1 Oxford St, Cambridge, MA 02138}

\keywords{Black holes, Absorption, LGT}
\pdfsubject{Black holes, Absorption, LGT}


\makeatletter
\thispagestyle{empty}

\begin{flushright}
\begingroup\ttfamily\@preprint\par\endgroup
\end{flushright}

\begin{centering}
\begingroup\Large\normalfont\bfseries\@title\par\endgroup
\vspace{1cm}

\begingroup\@authors\par\endgroup
\vspace{5mm}

\begingroup\itshape\@affiliationone\par\endgroup
\vspace{3mm}
\begingroup\itshape\@affiliationtwo\par\endgroup
\vspace{3mm}

\begingroup\ttfamily\@email\par\endgroup
\vspace{0.25cm}

\begin{minipage}{13cm}
 \begin{abstract}
The absorption rate of low-energy, or \emph{soft}, electromagnetic radiation by spherically symmetric black holes in arbitrary dimensions is shown to be fixed by conservation of energy and large gauge transformations. We interpret this result as the explicit realization of the Hawking-Perry-Strominger Ward identity for large gauge transformations in the background of a non-evaporating black hole. Along the way we rederive and extend previous analytic results regarding the absorption rate for the minimal scalar and the photon.
 \end{abstract}
\end{minipage}

\vspace{3mm}
\rule{\textwidth}{.5mm}
\vspace{-1cm}

\end{centering}

\makeatother

\tableofcontents
\newpage

\section{Introduction}

Recently, a number of intriguing connections have been made between
three physical ideas: ``large gauge transformations'', ``memory''
effects~\cite{zel1974radiation,braginsky1979electromagnetic,christodoulou1991nonlinear}, and soft theorems (e.g., in \cite{weinberg1996quantum}). Beyond showing old
results~\cite{Kapec:2015ena,He:2014cra,Kapec:2014zla,Lysov:2014csa} are the result of underlying symmetry principles,
these results have motivated new investigations into soft scattering
and memory~\cite{Strominger:2014pwa}. This paper extends this ``triangle'' of related
phenomena into a ``square''; the new vertex being low energy
absorption rates.

The basic insight is that local transformations that have nonvanishing
support on the boundary of spacetime need not be
gauged,\footnote{Here, ``gauged'' is used to mean unphysical degrees
  of freedom that are modded out of the Hilbert space. This, in of
  itself, is not new e.g.~\cite{Brown:1986nw}; what is new is the
  realization that there are many more interesting transformations
  than previously realized in asymptotically flat spacetime. The
  specific transformations are interesting because the corresponding
  longitudinal modes are, in fact, the low energy limit of the usual
  transverse modes.} in which case they are physically relevant
symmetries with all that entails: conserved currents, charges, and
Ward identities. The transformations are energy-preserving \emph{shift
  symmetries}: inhomogeneous transformations of the field. Shift
symmetries suggest spontaneous symmetry breaking, and indeed one can
interpret the transformations that are not isometries of the vacuum as
spontaneously broken symmetry generators. With new conserved charges
in theories of gravity, it is natural to ask, as
Hawking--Perry--Strominger~(HPS) did~\cite{Hawking:2016msc}, about the
implications for black hole evaporation.

While the exact role of these large
gauge charges in the black hole information problem has not been fully understood as of yet, we show
that shift symmetries of the above kind constrain the form of low
energy black hole absorption. Indeed, we show that conservation of
energy along with large gauge symmetry conservation laws fix the leading
low-energy photon absorption rate of spherically symmetric black holes
in $(p+2)$-dimensions. Beyond the conceptual advance in clarifying the
role of large gauge transformations for black hole physics, and in
demonstrating that low energy absorption follows from symmetry
arguments, our result for absorption of angular momentum $\ell$
electromagnetic waves for general $p$ and general charge $Q$ black
holes in asymptotically flat space appears to be a new result in the
literature.\footnote{The four-dimensional Reissner--Nordstr\"{o}m
  result appears in~\cite{Fabbri:1975sa, Fabbri:1977ux}; the general
  $p$ result for Schwarzschild in~\cite{Crispino:2000jx}; and for extremal charge
  in~\cite{Crispino:2010fd}. Note that an even more general result for Kerr solutions can be found in \cite{Page:1977um}.} Insofar as Weinberg's soft theorem is equivalent to the
Ward identity for large gauge transformations, one might state our
basic result as ``Weinberg's soft theorem fixes the leading low energy
black hole absorption rate''; however, we prefer to say that (large)
gauge symmetry implies both Weinberg's soft theorem and low energy
black hole absorption.

This paper serves as an illustration of the basic idea. The approach
can obviously be straightforwardly generalized in a number of
directions. The liminal arguments presented here apply to any other
fields with inhomogeneous symmetry transformations---shift
symmetries---including gravitons and gravitinos,
cf.~\cite{Avery:2015iix, Lysov:2015jrs}. One should be able to
generalize our results to spinning black holes and black branes, as
well. Minor modifications should allow one to apply our methods to
asymptotically anti-de Sitter spacetimes, which should give low energy
transport coefficients. Let us further note that the conservation laws
we write down are valid quite generally, even in spacetimes without an
event horizon.

We present the calculation as a classical scattering problem. This
serves two purposes: first, it agrees better with the existing black
hole scattering literature and second, it may make the calculation
accessible to a broader audience. The latter is particularly
important, given confusion observed by the authors regarding the
significance of HPS \cite{Mirbabayi:2016axw,Sheikh-Jabbari:2016npa} and more generally the connection between large
gauge transformations and infrared physics \cite{Gabai:2016kuf,Gomez:2016hxz}.

Our paper is organized as follows. In the next section, we 
present the conventions we be use in the rest of the paper. In
sec.~\ref{sec:minimal-scalar}, we derive the general result for
minimal scalar found in~\cite{Das:1996we} using only conservation of
energy and the shift symmetry of the scalar. We show that these two
symmetries fix the absorption of low-energy scalar waves uniquely. The
constant shift symmetry of the minimal scalar serves as a toy version
of the electromagnetic gauge symmetry. Many of the equations and much of the
reasoning carry over to the electromagnetic case with only small
changes. We then expand this analysis to the photon in
sec.~\ref{sec:photon}. Finally, we conclude with a discussion of the
results and an outlook describing future work. In the appendices, we
collect some useful results for reference.

\section{Conventions}
\label{sec:conventions}

Following Das--Gibbons--Mathur~(DGM)~\cite{Das:1996we}, we work with
$(p+2)$-dimensional spherically symmetric black hole spacetime metrics
of the form
\begin{equation}\label{eq:metric}
ds^2 = -f(r) dt^2 + g(r)\big(dr^2 + r^2d\Omega_p^2\big),
\end{equation}
with horizon at radius $r=r_H$ which is determined by
$f(r_{H})=0$. The squared line element on the unit $p$-sphere is denoted by $d\Omega_{p}^{2}$. The area of the horizon is given by\footnote{We correct
  a minor typo in~\cite{Das:1996we}.}
\begin{equation}
A_H = \big(r_H^2 g(r_H)\big)^\frac{p}{2} \omega_p = R_H^p \omega_p\qquad
\omega_p = \frac{2\pi^\frac{p+1}{2}}{\Gamma(\frac{p+1}{2})},
\end{equation}
where $\omega_p$ is the volume of the unit $p$-sphere and $R_H$ is the
normalized radius of the sphere, defined from the above by
$R_{H}^{2} = r_{H}^{2}g(r_{H})$. The functions $f(r)$ and $g(r)$ are
given for the Schwarzschild black hole and Reissner--Nordstr\"om (RN) black
holes in Appendix~\ref{sec:schw-rn-metr}. The functions $g(r)$ and
$f(r)$ have particularly nice properties for our calculation. The
former is finite at the horizon of the black hole while the latter is
of order ${\cal O}(r-r_{H})^{2}$ at the horizon. Note these
coordinates are valid outside the horizon, at $r > r_H$. For the extremal RN black hole, the horizon gets mapped to $r_{H}=0$
and the function $g(r)$ diverges at the horizon in this case. The
exact expressions for $p+2$ dimensions and more details can be
found in appx~\ref{sec:schw-rn-metr}.

\section{Minimal Scalar}
\label{sec:minimal-scalar}

The minimal scalar enjoys a shift symmetry $\phi \to \phi + \epsilon$ that one can think of as a
toy version of the large gauge transformations to be considered in the
sequel. While the low energy $\ell = 0$ absorption result we calculate
already exists in~\cite{Das:1996we}, the connection to this symmetry
has not been emphasized.\footnote{Although the resulting conservation laws do appear in~\cite{Paulos:2009yk,Iqbal:2008by}.}

Begin by considering a massless, minimally coupled scalar $\phi$ in an arbitrary curved background:
\begin{equation}
S = \int \drm^{p+2}x\sqrt{-g}\, \tfrac{1}{2}g^{\mu\nu}\pd_\mu\phi\pd_\nu\phi.
\end{equation}
For the metric in~\eqref{eq:metric}, the equations of motion take the
form
\begin{equation}
E(\phi) = -\frac{1}{f(r)}\pd_t^2\phi 
 + \frac{1}{r^p \sqrt{f\, g^{p+1}}}\pd_r\left(r^p \sqrt{f\, g^{p-1}}\pd_r\phi\right)
 + \frac{1}{g(r) r^2}\hat{\Delta}_p\phi,
\end{equation}
where $\hat{\Delta}_p$ is the Laplacian on the unit $p$-sphere. Let us
use the time translation and spherical symmetry of the background
to decompose $\phi(x)$ into modes:
\begin{equation}
\phi(t, r, \Omega) = \sum_{\ell, m}\int\frac{\drm \omega}{2\pi} \phi_{\omega, \ell, m}(r)\, Y_{\ell, m}(\Omega)e^{-i\omega t}.
\end{equation}
In the following, we will perform our analysis on a $\ell=0$ fixed $\omega$ mode, frequently
suppressing the mode labels to avoid notational clutter.

\subsection{Symmetries}

We will use two symmetry properties of the scalar field. First of all, the rigid shift symmetry with transformation
\begin{equation}
\phi(x) \mapsto \phi(x) + \epsilon,
\end{equation}
where $\partial_\mu\epsilon = 0$. The corresponding conserved current
is given by
\begin{equation}\label{eq:shift-current}
j^\mu = \epsilon g^{\mu\nu}\pd_\nu \phi,
\end{equation}
which almost trivially satisfies
\begin{equation}
\nabla_\mu j^\mu = E(\phi)\delta_\epsilon \phi,
\end{equation}
as required. This is the statement of conservation of canonical momentum in~\cite{Paulos:2009yk,Iqbal:2008by}. Additionally, we will need conservation of energy. For this we need the stress tensor, which is given by
\begin{equation}
T^{\mu\nu} = \frac{2}{\sqrt{-g}}\frac{\delta S}{\delta g_{\mu\nu}} 
         = \nabla^\mu\phi\nabla^\nu\phi - \frac{1}{2}g^{\mu\nu}g^{\rho\sigma}\nabla_\rho\phi\nabla_\sigma\phi
\end{equation}
It is possible to express $T$ in terms of $j$, but the conservation law is independent of the conservation of $j$. As our metric has a timelike
Killing vector, $\xi = \pd_t$, conservation of energy
follows
\begin{equation}\label{eq:scalar-energy-curr}
\nabla_\mu T^{\mu 0} = \nabla_\mu( T^{\mu\nu}\xi_\nu) = 0.
\end{equation}

\subsection{Solutions}

In order to find the $\ell = 0$ low energy absorption rate, we need to
solve the $\ell = 0$ equation of motion in two limits: the
asymptotically flat region, $r\gg r_H$, where $f, g \to 1$; and the
near-horizon region, $(r-r_H) \ll R_H$. The conservation laws relate
the two regions' small $\omega$ behavior without having to say
anything about the interior. Note that solving in these two regions is
necessary, anyway, to define what we mean by the absorption rate, and
to impose physically appropriate boundary conditions at the horizon.

\paragraph{Asymptotically Flat Limit}
For $r\gg r_H$, $f(r)=g(r)=1$ and we just have the flat equations of
motion.  For $\ell = 0$, the solution is
\begin{equation}
\phi_\text{flat}(r) = (\omega r)^{-\frac{p-1}{2}}
  \left(A_\omega\, J_\frac{p-1}{2}(\omega r) + B_\omega\, Y_\frac{p-1}{2}(\omega r)\right)\qquad r\gg r_H,
\end{equation}
for $p> 1$. We take two further limits of the above result. First,
taking $\omega r \gg 1$, we may use the Bessel functions' large
argument asymptotic form to write
\begin{equation}\label{eq:phi-sin-cos}
\phi_\text{flat}(r) \simeq \frac{2}{\sqrt{2\pi(\omega r)^p}}\left(
   A_\omega \cos\big(\omega r - \tfrac{(p-2)\pi}{4}\big) 
 + B_\omega \sin\big(\omega r - \tfrac{(p-2)\pi}{4}\big)\right)\qquad \omega r \gg 1.
\end{equation}
Second, taking $\omega r \ll 1$ (but keeping $r\gg r_H$) we may use
the small argument limit of the Bessel functions to write
\begin{equation}\label{eq:phi-powers}
\phi_\text{flat} \simeq \frac{1}{2^{\frac{p-1}{2}}}\left( \frac{A_\omega}{\Gamma(\frac{p+1}{2})}
- \frac{B_\omega\Gamma(\frac{p-1}{2})}{\pi}\left(\frac{\omega r}{2}\right)^{-(p-1)}
\right)
\qquad \omega r \ll 1.
\end{equation}
It is the above limiting form which gets related to the near-horizon
physics by the two conservation laws.

\paragraph{Near-Horizon Limit}
For the near-horizon analysis, it is convenient to define a new radial
coordinate $\rho$ ($\tau$ in \cite{Das:1996we}) such that
\begin{equation}
\pd_\rho = r^p\sqrt{f(r)\, g(r)^{p-1}}\pd_r.
\end{equation}
With this radial coordinate, the equation of motion takes the form
\begin{equation}
E(\phi) = \frac{1}{f(r)\big(r^2g(r)\big)^p}\left[
\pd_\rho^2 - \big(r^2g(r)\big)^p\pd_t^2 + f(r) \big(r^2 g(r)\big)^{p-1}\Delta_p\right]\phi.
\end{equation}
At the horizon of a nonextremal black hole, $f$ has a double zero and
$g$ is regular. The $\ell = 0$ mode equation for $r \sim r_H$ becomes
\begin{equation}
\big(\pd_\rho^2 + R_H^{2p}\omega^2\big)\phi_\omega(\rho) = 0,
\end{equation}
with solutions $\exp(\pm i\omega R_H^p \rho)$. Ingoing boundary
conditions choose the negative sign. Thus, for small $r$ the solution
takes the form
\begin{equation}
\phi_\omega(r) \simeq \phi_H\, e^{-i\omega R_H^p \rho}.
\end{equation}

\paragraph{Absorption Rate} We can use the asymptotically flat
solutions to give the absorption rate in terms of $A_\omega$ and
$B_\omega$. Explicitly, from eq.~\eqref{eq:phi-sin-cos} the
absorption rate is given by
\begin{equation}\label{eq:scalar-rate}
\Gamma_\text{abs} = 1 - 
  \left|\frac{1 - i\frac{B_\omega}{A_\omega}}{1+i\frac{B_\omega}{A_\omega}}\right|^2.
\end{equation}
Instead of working with the coefficients $A$ and $B$, let us rewrite
 the ratio $B/A$ in terms of $\phi'(R)/\phi(R)$ for $R \gg r_H$ and
$\omega R \ll 1$. Using~\eqref{eq:phi-powers}, we may write
\begin{equation}\label{eq:B-by-A}
\frac{B_\omega}{A_\omega} = \left(\frac{\omega R}{2}\right)^{p-1}
\frac{\pi(p-1)}{2\,\Gamma(\frac{p+1}{2})^2}
\frac{R\frac{\phi'(R)}{\phi(R)}}{p-1+R\frac{\phi'(R)}{\phi(R)}},
\end{equation}
after dropping subleading terms in $\omega R$. Below we show that
conservation of energy and ``canonical momentum'' fixes
$\phi'(R)/\phi(R)$, for $R\gg r_H$ and $\omega R \ll 1$, in terms of
$\phi'(r_H)/\phi(r_H)$. From the near-horizon solution, we see that
\begin{equation}\label{eq:phip-by-phi}
\frac{\phi'(r_H)}{\phi(r_H)} = -i\omega \sqrt{\frac{g(r_H)}{f(r_H)}},
\end{equation}
where there is an implicit regulator on $r_H$, since $f(r)$ has a
double zero at the horizon. We will see that the regulator cancels out of the
absorption rate.

\subsection{Conservation laws}

We have two conserved currents, eqs.~\eqref{eq:shift-current}
and~\eqref{eq:scalar-energy-curr}, corresponding to shift symmetry and
time translation symmetry. The shift symmetry implies the existence of
a constant mode, which we would like to relate to the $\omega \to 0$
limit of the $\ell=0$ ($s$-wave) mode. The conservation laws imply that for any subregion $R$,
\begin{equation}
\label{eq:2}
\oint_{\pd R} \star j = \int_R d\star j = \int_R \bm{\epsilon} E(\phi)\delta\phi \wkeq 0,
\end{equation}
where we use the notation of~\cite{Avery:2015rga} to emphasize that the last
equality follows only after using equations of motion. It is up to us
to find a convenient contour $C=\pd R$ that usefully constrains the
dynamics under consideration. To follow HPS, it would seem natural to
choose the spacetime outside the black hole event horizon; however, we
need to regulate the contour. For the steady state absorption process
under consideration, the following regulated contour seems most
convenient:
\begin{equation}
C = \Sigma_- \cup \Lambda \cup \Sigma_+ \cup H,
\end{equation}
with
\begin{equation}\begin{aligned}\label{eq:sym-box-contour}
\Lambda&: r=R\quad t\in [-T, T]\\
\Sigma_-&:  t = -T\quad r\in [r_H + \delta, R]\\
\Sigma_+&:  t = T\quad r\in [r_H + \delta, R]\\
H&: r=r_H + \delta \quad t\in [-T, T].
\end{aligned}.
\end{equation}

\begin{figure}[htpb]
  \centering
  \includegraphics[width = .5\textwidth]{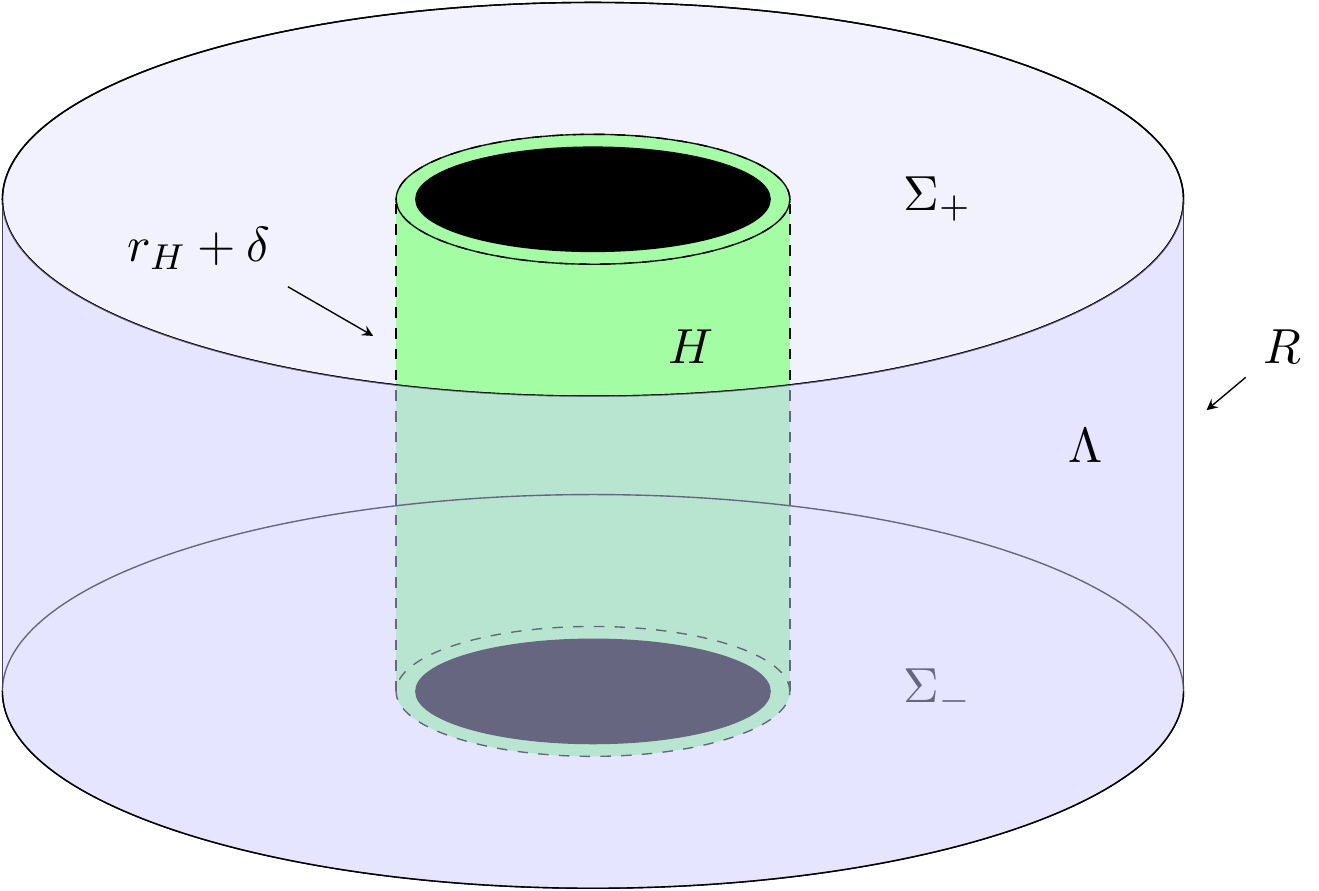}
  \caption{\small A cartoon of the integration regions. Time runs upward and each $S^1$ of the cylinder represents an $S^p$. The top and bottom of the cylinder are at $\pm T$, the respective regions $\Sigma_{\pm}$ are the spatial integration regions. $\Lambda$ is blue, $H$ is green. At a distance of $\delta\ll R_H$ to the horizon of the black hole, located at $r = r_{H}$, we put a \emph{stretched horizon}. Similarly, at $R\gg r_{H}$ we put a \emph{stretched infinity}. At the end of the calculation we will remove both regulators at the same time.}
  \label{fig:1}
\end{figure}

See Fig.~\ref{fig:1} for a depiction of the surfaces.
Note that we are not working with null surfaces, which is different
from HPS; although this should not change the physics. In the limit
$\delta \to 0$ and $R, T\to \infty$ we should cover the exterior
spacetime. More specifically we shall take $\delta \to 0$ first, and
$T$ and $R$ large with $T \gg R $.

Using the shift current and the contour $C$, \eqref{eq:2} becomes 
\begin{multline}
\int_{r_H}^R \drm r\, r^p\sqrt{\frac{g(r)^{p+1}}{f(r)}}\int_{0}^\infty\drm \omega
   \frac{\sin(\omega T)}{\pi\omega}\omega^2(\phi_\omega(r) - \phi_{-\omega}(r))\\
 = R^p\sqrt{f(R) g(R)^{p-1}}\int_{-\infty}^\infty\drm \omega\frac{\sin(\omega T)}{\pi\omega}
   \phi'_\omega(R) 
  - r_H^p\sqrt{f(r_H) g(r_H)^{p-1}}\int_{-\infty}^\infty\drm \omega\frac{\sin(\omega T)}{\pi\omega}
   \phi'_\omega(r_H).
\end{multline}
Recalling that
\begin{equation}
\lim_{T\to\infty} \frac{\sin (\omega T)}{\pi\omega} = \delta(\omega),
\end{equation}
we find in the small $\omega$ limit the spacelike portions of the
contour are subleading and thus
\begin{equation}\label{eq:shift-conservation-law}
R^p\sqrt{f(R) g(R)^{p-1}} \phi'_\omega(R) \simeq r_H^p\sqrt{f(r_H) g(r_H)^{p-1}}\phi'_\omega(r_H) 
  + O(\omega^2).
\end{equation}
As we noted before, this result can be found in~\cite{Paulos:2009yk,Iqbal:2008by} as conservation of
canonical momentum. Notice that the conservation law is a useful
constraint, precisely because it is indifferent to the behavior of
$\phi$ (or equivalently the metric coefficients $f$ and $g$) in the
middle region away from the horizon and asymptotically flat region.

The shift conservation law~\eqref{eq:shift-conservation-law} is not
sufficient to determine the absorption rate. We need two conditions
relating the near-horizon mode behavior to the asymptotic mode
behavior. Thus, let us turn to conservation of energy.

Let us take the same contours as in~\eqref{eq:sym-box-contour}. Note
that there is an important difference from the shift symmetry: all of
the spherical modes contribute to the energy. Let us work on the
spacelike slices first:
\begin{equation}
\int_{\Sigma}\star J = -\frac{1}{2}\int\drm \Omega\, r^p\sqrt{f(r)g(r)^{p-1}}\phi\phi'\Big|_{r_H}^R
-\frac{1}{2}\int\drm \Omega\int\drm r\, r^p\sqrt{f g^{p+1}}\left[\frac{\dot{\phi}^2}{f}
      - \frac{\phi\ddot{\phi}}{f} - \phi\, E(\phi) \right].
\end{equation}
Note that the second term is $O(\omega)$, after using the equations
of motion.  The spacelike integrals again give subleading corrections to the
leading result when one takes $T\to\infty$. The timelike integrals
give
\begin{equation}
\int_{\Lambda +H}\star J = \omega_p\int\frac{\drm \omega_1}{2\pi}\int\drm \omega_2\,
  \frac{\sin((\omega_1 + \omega_2)T)}{\pi(\omega_1+\omega_2)}(-i\omega_2)\,
  \left[r^p\sqrt{f(r)g(r)^{p-1}}\phi'_{\omega_1}(r)\phi_{\omega_2}(r)\right]_{r_H}^R.
\end{equation}
In the large $T$ limit, the $\sin$ factor gives a
$\delta(\omega_1+\omega_2)$, and we are left with
\begin{equation}
R^p\sqrt{f(R)g(R)^{p-1}}\phi_{-\omega}(R)\phi_\omega'(R) = r_H^p\sqrt{f(r_H)g(r_H)^{p-1}}\phi_{-\omega}(r_H)\phi_\omega'(r_H).
\end{equation}
Combining conservation of energy with the shift conservation law we
find that $\phi_\omega$ approaches the constant shift mode in the
small $\omega$ limit:\footnote{We assume a single coherent $\omega$ is
  excited. In general, there is a sum over all $\omega$; each
  $\pm\omega$ contributes to the energy.}
\begin{equation}
\phi_\omega(r_H) \simeq \phi_\omega(R) + O(\omega),
\end{equation}
and one can find the identity
\begin{equation}
R^p\sqrt{f(R) g(R)^{p-1}}\frac{\phi'(R)}{\phi(R)} = r_H^p\sqrt{f(r_H)g(r_H)^{p-1}}\frac{\phi'(r_H)}{\phi(r_H)}.
\end{equation}
Using that $f(R), g(R) \to 1$ for large $R$ and~\eqref{eq:phip-by-phi}, this becomes
\begin{equation}
R\frac{\phi'(R)}{\phi(R)} = -i\omega R\, \left(\frac{R_H}{R}\right)^p,
\end{equation}
which we can plug into~\eqref{eq:B-by-A} and~\eqref{eq:scalar-rate} to
find the absorption rate. In the limits we are working, one finds
\begin{equation}
\Gamma_\text{abs} = 4\frac{\pi}{\Gamma(\frac{p+1}{2})^2} \left(\frac{\omega R_H}{2}\right)^p,
\end{equation}
which agrees with Ref.~\cite{Das:1996we}, after using standard Gamma
function identities. Note that the dependence on all of the regulators
has dropped out. Additionally, the metric dependence has also dropped out. This is a well-known result: The absorption rate for minimal scalars only depends on the area $A_{H}$ of the black hole horizon, but is otherwise independent of the function $g(r)$. We will see that this is not true for electromagnetic radiation.

\section{Photon}
\label{sec:photon}

We now turn to the photon. In the last section, we used the shift
symmetry of the massless scalar to derive our results. In the present
case, the shift symmetry is replaced by the so called (large) gauge
symmetry. That is, transformations of the photon field $A_{\mu}$
\[
A_{\mu}\to A_{\mu} + \del_{\mu} \lambda(x),
\]
where $\lambda(x)$ is a function of the coordinates on the sphere
$\theta^{A}$ at asymptotic infinity, i.e.,
\[
\lim_{r\to\infty}\lambda(t,r, \theta^{A}) = \lambda(\theta^{A}).
\]
These are the so called large gauge transformations defined
in~\cite{Strominger:2013lka,He:2014cra,Kapec:2015ena}. To fix our conventions, on a curved
background with metric $g_{\mu\nu}(x)$, we take the Lagrangian density
for the photon to be\footnote{Note that we do not include a dynamical metric, so that we are not allowing graviton--photon mixing, however, this seems to be a subleading effect in the small $\omega$ limit.}
\begin{equation}
\mathscr{L} = -\frac{\sqrt{-g}}{4} F_{\mu\nu}F^{\mu\nu},
\end{equation}
which implies the equations of motion
\begin{equation}
E^\nu = -\nabla_\mu F^{\mu\nu}= - \Box A_\nu + \nabla_\nu (\nabla^\mu A_\mu) + {R^\lambda}_\nu A_\lambda.
\end{equation}
The Noether current for gauge transformations---the derivation can be
found in, e.g.,~\cite{Avery:2015rga}---is given by
\begin{equation}
j^\mu = F^{\mu\nu}\pd_\nu\lambda.
\end{equation}
Note that the Noether current has similarities with the shift
current~\eqref{eq:shift-current}; it depends on the field strength $F_{\mu\nu}$ and is accompanied by the transformation
parameter $\lambda(x)$. To draw a parallel with the minimal scalar let us note that the time component of $j^{\mu}$---which we ultimately use to determine a conserved quantity for this current---only depends on the canonical momentum density $F_{0i}=E_{i}$ of the field $A_{i}$. We may again speak about the conservation of canonical momentum. However, there are also glaring
dissimilarities. The new ingredient here is that $\lambda$ is a
function on the sphere with a clearly determined boundary value at
$r\to\infty$. Apart from that, $\lambda$ does not, \emph{a priori},
satisfy any constraints, so in the bulk of the space time,
$\lambda(x)$ is undetermined. The current is conserved for all
$\lambda$, so that it is up to us to determine the gauge
transformations that give useful conservation laws.  We will find
appropriate constraints for it in due course. In particular, we will
show that the conservation law is useful when $\lambda$ solves the
$\omega = 0$ photon equation with appropriate fall off for large $r$. 
It is clear however,
that, since $\lambda$ is now a function of at least the coordinates on
the sphere, we should be able to get more than just a single $\ell$ mode
of the field $A_{\mu}$. Indeed, we see that conservation of canonical momentum for the minimal scalar becomes conservation of canonical momentum \emph{density} for the photon; there is now a zero mode for every $\ell>0$. The on-shell conservation of $j^{\mu}$ follows
directly
\begin{equation}
\nabla_\mu j^\mu = -E^\nu\nabla_\nu\lambda + \frac{1}{2}F^{\mu\nu}[\nabla_\mu, \nabla_\nu]\lambda 
  = -E^\nu\nabla_\nu\lambda.
\end{equation}
As with the minimal scalar, we also need to give the stress tensor for
the photon field in curved spacetime. It takes the form
\begin{equation}
T_{\mu\nu} = F_{\mu\rho}{F^\rho}_\nu + \frac{1}{4}g_{\mu\nu}F_{\rho\sigma}F^{\rho\sigma}.
\end{equation}
Its conservation follows from
\begin{equation}
\nabla^\mu T_{\mu\nu} 
  = \frac{1}{2}E^\rho F_{\rho\nu}
   - \frac{1}{2}F^{\rho\sigma}({R^\lambda}_{\nu\rho\sigma} 
     + {R^\lambda}_{\sigma\nu\rho} + {R^\lambda}_{\rho\sigma\nu})A_\lambda,
\end{equation}
where the second term vanishes due to the first Bianchi identity for the Riemann
tensor.

To proceed, we shall write down a mode expansion for $A_{\mu}$. As the
background \eqref{eq:metric} is spherically symmetric, we would like
to exploit this circumstance by labeling modes with the following
commuting set of operators:\footnote{NB: These operators commute with
  each other and the equations of motion.} since the spacetime is
static and spherically symmetric, we have a timelike Killing vector
$\del_{t}$ and rotational symmetry $SO(p+1)$. Thus we can use the
Hamiltonian $H = i\mathcal{L}_{\pd_t}$ and the $SO(p+1)$ Casimir
$J^2$. In four dimensions ($p=2$), we also include the parity
transformation $P$. The parity is relevant in this special case, as it
allows us to distinguish the spherical harmonics $\hat{\nabla}Y_{lm}$
from $\hat{W}_A = \hat{\epsilon}_{AB}\hat{\nabla}^B Y_{lm}$. In higher
dimensions, the analogous two modes have different $J^2$. For
convenience, let us decompose the Lorentz vector index $\mu$ into
$i, j, k \in (t, r)$ and $A, B, C\in \Omega$. The $SO(p+1)$ Casimir
acts as
\begin{equation}
J^2 A_j = -\hat{\Delta} A_j\qquad J^2A_A = -(\hat{\Delta} - p+1)A_A.
\end{equation}
If we let $m$ represent the set of eigenvalue(s) of the generators of
the Cartan subalgebra of $SO(p+1)$ we may symbolically write even
higher dimensional spherical harmonics as $Y_{\ell m}$.  Notice that
$m$ is a single number only for $p=2$. As we will only use a very
limited set of properties of the spherical harmonics on $S^{p}$, we
will not give many details about these functions. More details can be
found in~\cite{Chodos:1983zi, Rubin:1983be}. For us, the relevant
information is the following. On each $S^p$, there are two kinds of
vector spherical harmonics: there is a \emph{gradient mode},
$\hat{\nabla}_A Y$, and there is a \emph{divergence-free mode}, $\hat{W}_A$,
with $\hat{\nabla}^A\hat{W}_A = 0$~\cite{Chodos:1983zi,
  Rubin:1983be}. These modes have eigenvalues
\begin{equation}
\hat{\Delta} (\hat{\nabla}_A Y_\ell) = -\big[\ell(\ell+p-1)-(p-1)\big](\hat{\nabla}_A Y_\ell)\qquad
\hat{\Delta}\, \hat{W}^{(\ell)}_A = -\big[\ell(\ell+p-1)-1\big]\hat{W}^{(\ell)}_A,
\end{equation}
and thus
\begin{equation}\label{eq:orthogonalSH}
J^2(\hat{\nabla}_A Y_\ell) = \ell(\ell+p-1)\hat{\nabla}_A Y_\ell\qquad
J^2\hat{W}^{(\ell)}_A = [\ell(\ell+p-1)+p-2] \hat{W}^{(\ell)}_A.
\end{equation}
Note that it is only for $p=2$ that the two Casimir eigenvalues
coincide. The degeneracies of the gradient and divergence-free mode
are given by
\begin{equation}
d_s(p,\ell) = \frac{(\ell+p-2)!}{(p-1)!\,\ell!}(2\ell+p-1)\qquad
d_V(p, \ell) = \frac{\ell(\ell+p-1)\,(2\ell+p-1)(\ell+p-3)!}{(p-2)!\,(\ell-1)!},
\end{equation}
respectively. From \eqref{eq:orthogonalSH} it is quite obvious that
the $\hat{W}$ cannot mix with the scalar modes since they have a
different value for the Casimir for $p>2$. For the special case of
$p=2$ we use---as we explained above---parity to distinguish these
two modes.

To proceed, we define a ``gradient mode''
\begin{equation}
A^{(+)}_\mu = \big(A(r) Y_{\ell,m}(\Omega), B(r) Y_{\ell, m}(\Omega), C(r) \hat{\nabla}_AY_{\ell,m}\big)e^{-i\omega t},
\end{equation}
as well as a ``solenoidal mode'' which is given by
\begin{equation}
A^{(-)}_\mu = \big(0, 0,D(r) \hat{W}^{(\ell)}_A(\Omega)\big)e^{-i\omega t}.
\end{equation}
For convenience, throughout the rest of the text, we may refer to these as
$+$ and $-$ modes, respectively. In the following, we focus on the
gradient or $+$ mode and give only a few comments on the solenoidal
mode.

Let us also define a ``longitudinal'' or gauge mode:
\begin{equation}
L_\mu = \partial_\mu(\Lambda(r) Y_{lm}(\Omega) e^{-i\omega t}) 
     = \big(-i\omega\Lambda(r) Y_{lm}(\Omega), \Lambda'(r) Y_{lm}(\Omega), 
            \Lambda(r) \hat{\nabla}_A Y_{lm}(\Omega)\big)e^{-i\omega t}.
\end{equation}
For $\omega > 0$ this mode decouples and is unphysical, as expected;
however, in the $\omega \to 0$ limit the longitudinal and gradient
modes degenerate as we see below explicitly in temporal gauge.

\subsection{Gauge choice and radial equations of motion}

So far, we haven't chosen a gauge. We shall remedy this situation
presently. It turns out that the calculation is particularly
convenient in temporal gauge
\[
A_t = 0.
\]
When imposing this gauge on the $+$ mode defined above we find that
$A=0$ and $B$ is related to $C'(r)$ via the Gauss constraint, while the
$-$ mode already satisfies the gauge condition. The $+$ mode then
takes the form
\begin{equation}
A^{+}_\mu = \left(0, 
  C'(r)\frac{\ell(\ell + p-1) f(r)}{\ell(\ell + p -1) f(r)-\omega^2 r^2 g(r)}Y_{lm},C(r)\hat{\nabla}_A Y_{lm}\right)_\mu e^{-i\omega t}.
\end{equation}
The rest of the equations of motion imply that this mode is a solution
provided $C(r)$ satisfies the radial equation of motion
\[\label{eq:C-DE}
C''(r) + \frac{\drm}{\drm r}\log\left[\frac{r^p\sqrt{f(r) g(r)^{p-1}}}{\ell(\ell+p-1)f(r)-\omega^2 r^2g(r)}\right] C'(r) + \left(\omega^2 \frac{g(r)}{f(r)} - \frac{\ell(\ell+p-1)}{r^2}\right)C(r) = 0.
\]
We may also calculate the field strength $F_{\mu\nu}$, which we 
use extensively in the calculation below. It is given by
\begin{equation}\begin{aligned}
\label{eq:goodsolution}
F_{tr} &= i\omega\frac{\ell (\ell+p-1) f(r)}{\omega^2 r^2 g(r)-\ell (\ell+p-1) f(r)}
  C'(r)\, Y_{\ell m} e^{-i \omega t}\\
F_{t A} &= -i \omega  C(r)\, (\hat{\nabla}_AY_{\ell m})e^{-i\omega t}\\
F_{r A} &= \frac{\omega^2r^2g(r)}{\omega^2r^2 g(r)-\ell(\ell+p-1)f(r)}C'(r)\,(\hat{\nabla}_A Y_{\ell m})e^{-i\omega t}\\
F_{A B} &= 0
\end{aligned}.
\end{equation} 
We put emphasis on this mode over the solenoidal mode as it is the one
which degenerates with the pure gauge mode $\lambda(x)$. Information
regarding the solenoidal mode can be found in
appx~\ref{sec:solenoidal-mode}. Observe that both the gradient and
solenoidal mode start with $\ell = 1$, not $\ell=0$, since the photon
has spin one.

As for $\lambda(x)$, after imposing the gauge condition, we find that
the residual gauge freedom is given by
$\lambda = \Lambda_{\ell m}(r)Y_{\ell m}(\Omega)$. This corresponds to a ``longitudinal
photon'' of the form
\begin{equation}
L_\mu = \partial_\mu \lambda = (0,\, \Lambda_{\ell m}'(r) Y_{\ell m}(\Omega),\, \Lambda_{\ell m}(r)\hat{\nabla}_AY_{\ell m}(\Omega)).
\end{equation}
Note that this is indeed the $\omega\to 0$ limit of $A^{(+)}_\mu$,
which suggests an identification $C_\omega \to \Lambda$ as
$\omega \to 0$. $C_\omega(r)$ solves a second order differential so
there are only two physically realized profiles of $r$ as
$\omega \to 0$. With this observation, one may correctly anticipate
that the interesting choice of $\Lambda_{\ell m}(r)$ should be one of those
modes.

\subsection{Solutions for the gradient mode}\label{sec:grad-mode-sols}

We are now in a position to study the solutions of eq.~\eqref{eq:C-DE}. It
turns out that it is sufficient to study the equation in \emph{three} limits. This is an important difference from the $\ell=0$
minimal scalar calculation, where we only needed two limits. We first
study the case $r\gg r_H$ where the functions $f(r)$ and $g(r)$ in
eq.~\eqref{eq:metric} go to $1$. Thus \eqref{eq:C-DE} turns into the
radial equation of motion for a photon propagating in flat space. The
second limit is the near horizon limit $r-r_{H}=\delta\ll
R_H$.
Finally, we investigate the equation \eqref{eq:C-DE} on the whole
space in the $\omega\to 0$ limit.\footnote{These three limits are
  essentially the same as the three regions considered in a matched
  asymptotic expansion approach, see e.g.~\cite{Fabbri:1975sa,
    Fabbri:1977ux}. The last case, the ``intermediate region'', is
  valid when
  $\omega^2 \frac{g(r)}{f(r)} \ll \frac{\ell(\ell + p-1)}{r^2}$; however, we use the $\omega =0$ solution for the gauge parameter, $\Lambda$, in our conservation law below.}

\paragraph{Asymptotically Flat Limit}
Let us first turn to the asymptotically flat case. We find that the equation of
motion for the function $C(r)$ becomes the second order ordinary differential equation
\begin{equation}\label{eq:Cp-flat}
C''(r) + \frac{(p-2)\omega^2 r^2 - p\,\ell(\ell+p-1)}{\omega^2 r^3 - \ell(\ell+ p -1)r}C'(r)
+ \left(\omega^2 - \frac{\ell(\ell+p-1)}{r^2}\right)C(r) = 0
\end{equation}
which has a general solution in terms of combinations of Bessel functions of the first kind $J_{\alpha}(r)$, and second kind $Y_{\alpha}(r)$. Explicitly, one finds the solution
\begin{equation}\label{eq:flatsoln}
C_{\rm flat}(r) = C_1\frac{\ell\, J_{\ell + \frac{p-1}{2}}(\omega r) - \omega r\, J_{\ell + \frac{p-3}{2}}(\omega r)}{(\omega r)^\frac{p-1}{2}}
 + C_2\frac{\ell\, Y_{\ell + \frac{p-1}{2}}(\omega r) - \omega r\, Y_{\ell + \frac{p-3}{2}}(\omega r)}{(\omega r)^\frac{p-1}{2}}.
\end{equation} for all $p>1$ and all $\ell>0$.

Just as for the scalar, two further limits are of interest. For
$\omega r \gg 1$, the solution takes the form
\begin{equation}\label{eq:leadflatlarge}
C_{\rm flat}(r) \simeq -   \frac{1}{(\omega r)^\frac{p-2}{2}}\left[
C_1 \sqrt{\frac{2}{\pi}} \cos\big(\omega r - (2\ell + p-2)\tfrac{\pi}{4}\big)
+ C_2 \sqrt{\frac{2}{\pi}} \sin\big(\omega r - (2\ell + p-2)\tfrac{\pi}{4}\big)\right].
\end{equation}
Conversely, when $\omega r\ll 1$, the solution is asymptotically and to leading order 
\[\label{eq:leadflatsmall} C_{\rm flat} = - C_{1} \frac{\ell+p-1}{2^{\frac{p-1}{2}}\Gamma\left(\ell + \frac{p+1}{2}\right)} \left(\frac{\omega r}{2}\right)^{\ell} - C_{2}\frac{\ell\ \Gamma\left(\ell + \frac{p-1}{2}\right)}{2^{\frac{p-1}{2}}\pi}\left(\frac{\omega r}{2}\right)^{-\ell-p+1}.\]

We return to~\eqref{eq:flatsoln} in the next section where we examine the conservation laws that lead to the absorption rate. Before we do that though, let us also examine the near-horizon limit of \eqref{eq:C-DE}.

\paragraph{Near-Horizon Limit}
In particular  we are interested in the near-horizon limit of \eqref{eq:C-DE} for very low frequency. Since the limits do not commute, let us fix a prescription to which we adhere in the rest of the text. In the following, we always take the near-horizon limit before we take the small $\omega$ limit, i.e., $r-r_{H}=\delta \ll \omega r_H^2$. Then we can rewrite the resulting differential equation in terms of a new radial coordinate $\rho(r)$ such that its derivative
\begin{equation}
\rho'(r) = \frac{\ell (\ell+p-1)f(r)-\omega^2 r^2g(r)}{r^p\sqrt{f(r)g(r)^{p-1}}},
\end{equation}
and the equation of motion becomes
\begin{equation}
C''(\rho) + \frac{(r^2g(r))^{p-2}}{\omega^2 - \frac{f(r)}{g(r)}\frac{\ell (\ell +p-1)}{r^2}}C(\rho) = 0.
\end{equation}

At the horizon of a nonextremal black hole, $f(r)$ has a double zero
and $g(r)$ is regular.\footnote{For extremal black holes, $f(r)$
  develops higher order zeroes and $g(r)$ goes to $1/f(r)$.} Thus we can
make an approximation for $f(r)$ in the near horizon limit where
$f(r_H+\delta) \approx \frac{1}{2}f''(r_H)\delta^2$ and
$g(r_H+\delta) \approx g(r_H)$. In the limit defined above,
$\delta \ll \omega r_H^2$ and $\ell $ fixed, the solution to the
resulting differential equation is
\begin{equation}\label{eq:nearhorsoln}
C(\rho) \simeq C_H e^{\pm\frac{i\, R_H^{p-2}}{\omega}\rho}.
\end{equation}
At the horizon of the black hole, we want to choose a solution which
is purely ingoing. We can do so by choosing the sign of the exponent
appropriately. Note that in the limit we are taking, the derivative of
our radial coordinate $\rho'(r) < 0$, which implies that increasing
$r$ is decreasing $\rho$. The ingoing solution is therefore actually
the solution with the positive sign in the exponent.

\paragraph{Zero-Energy Limit}
We now consider the zero energy limit. In order for the spacelike
contributions to drop out of the conservation law, we find that the
gauge parameter $\Lambda_{\ell m}(r)$ must satisfy this equation.  To derive
the solution of the equation of motion for $C(r)$ in the $\omega=0$
limit, we can employ a specific black hole background like
Schwarzschild or Reissner--Nordstr\"om and then infer the general
solution. The equation itself is
\[
C''(r) + \frac{\del}{\del r} \log\left(r^{p}\sqrt{\frac{g(r)^{p-1}}{f(r)}}\right)C'(r)-\frac{\ell(\ell + p -1)}{r^{2}}C(r)=0.
\] 
The Schwarzschild metric in higher dimensions can be found in appx~\ref{sec:schw-rn-metr}.

In the following, we use $r_{H}$ to rewrite the functions $f(r)$ and $g(r)$ in terms of dimensionless variables $x = \frac{r}{r_{H}}$. We insert the functions $f(r)$ and $g(r)$ for the Schwarzschild solution into the $\omega=0$ and get
\[ C''(x)+ \left(\frac{p}{x} - \frac{2 (p-1) \left(2 x^p-x\right)}{x^{2 p}-x^2}\right) C'(x)-\frac{\ell(\ell +p-1)}{x^{2}}C(x) =0.\] The solution to this equation is given in terms of hypergeometric ${}_{2}F_{1}$. Specifically, the two independent solutions are
\begin{align}\label{eq:zerosoln}
C_{\omega=0,p}^{(1)}(r) &= \left(\frac{r}{r_{H}}\right)^{\ell - p +1}\frac{1}{\sqrt{g(r)^{p-1}}}{}_{2}F_{1}\left(-\frac12,\frac{\ell}{p-1};\frac32 + \frac{\ell}{p-1};\left(\frac{r}{r_{H}}\right)^{2(p-1)}\right)\\
C_{\omega=0,p}^{(2)}(r) &= \left(\frac{r}{r_{H}}\right)^{-\ell - 2p +2}\frac{1}{\sqrt{g(r)^{p-1}}}{}_{2}F_{1}\left(-\frac12,-1-\frac{\ell}{p-1};\frac12 - \frac{\ell}{p-1};\left(\frac{r}{r_{H}}\right)^{2(p-1)}\right)
\end{align} where we already inserted the form of the general solution where the $g(r)$ parametrizes the dependence of the solution on the specific black hole background.

We need to investigate two limits of these solutions. The near-horizon limit of these functions is easily derived by noticing that the hypergeometric function needs to be evaluated at 1 where it is well known that
\[_{2}F_{1}(a,b;c;1) = \frac{\Gamma(c)\Gamma(c-a-b)}{\Gamma(c-a)\Gamma(c-b)}\] if the real parts $\Re(c)>\Re(a+b)$. This is fulfilled here for the first solution. The function in the second line goes to zero in this limit. Explicitly, the first function becomes
\[C_{\omega=0,p}(r_{H}) = \frac{2}{\sqrt{\pi g(r_{H})^{p-1}}}\frac{\Gamma\left(\frac{\ell}{p-1}+\frac32\right)}{\Gamma\left(\frac{\ell}{p-1}+2\right)}.\]
On the other hand, for large r, we want to pick out a solution which goes like $R^{-\ell-p+1}$. For large argument, the hypergeometric function satisfies
\[_{2}F_{1}(a,b;c;z) = \frac{\pi}{\sin \pi(b-a)}\left(\frac{(-z)^{-a}}{\Gamma(b)\Gamma(c-a)} -\frac{(-z)^{-b}}{\Gamma(a)\Gamma(c-b)}\right)\] so the coefficient of the second solution can be adjusted to cancel an unwanted contribution from the first solution. Then
\[C_{\omega=0,p}(R) = \left(\frac{R}{r_{H}}\right)^{-\ell-p+1}.\] Later, we will need the ratio of the near-horizon and the flat space value. It is (suppressing some labels on $C$)
\[\frac{C(R)}{C(r_{H})} = \frac{\sqrt{\pi}}{2}g(r_{H})^{-\frac{\ell}{2}}\left(\frac{R_{H}}{R}\right)^{\ell+p-1}\frac{\Gamma\left(\frac{\ell}{p-1}+2\right)}{\Gamma\left(\frac{\ell}{p-1}+\frac32\right)}\label{eq:zeroratio}\] where we again used the relation $R_{H}^{2} = g(r_{H})r_{H}^{2}$.

\paragraph{Absorption Rate}

With the calculations from the first two paragraphs, we can assemble the absorption rate as a function of the ratio of the coefficients $C_{i}$ of the flat space solutions \eqref{eq:flatsoln}. Explicitly, from \eqref{eq:leadflatlarge} it follows that the absorption rate is given by
\begin{equation}\label{eq:abs-21}
\Gamma_{\rm abs} = 1 - \left|\frac{1-i \frac{C_2}{C_1}}{1+i\frac{C_2}{C_1}}\right|^2.
\end{equation}
Below, in the calculation of the cross section, we will encounter two ratios which we shall give here for later reference. The first is a ratio for the near-horizon solutions
\begin{equation}\label{eq:nearratio}
\frac{C'(r_H)}{C(r_H)} = \frac{i\, R_H^{p-2}}{\omega}\rho'(r) = -i\omega\sqrt{\frac{g(r_H)}{f(r_H)}},
\end{equation}
where there is an implicit regulator on $r_H$, which cancels out of the absorption calculation.

Conversely, when we take the radial coordinate $r$ to be much larger than the radius of the black hole $r\to R\gg r_H$, it is appropriate to
use~\eqref{eq:Cp-flat}. Simultaneously taking $\omega R \ll 1$ allows us to use the small argument expansion of the Bessel functions \eqref{eq:leadflatsmall}. In this limit the following equation is applicable
\begin{equation}\label{eq:C2divC1p}
\frac{C_2}{C_1} = \left(\frac{\omega R}{2}\right)^{2\ell + p -1}
\frac{\pi(\ell + p -1)(2\ell + p-1)}{2\ell\,\Gamma(\ell + \frac{p+1}{2})^2}
\frac{\ell - R\frac{C'(R)}{C(R)}}{(\ell + p -1) + R\frac{C'(R)}{C(R)}}.
\end{equation}
Thus we can turn the absorption rate into a function of the ratio
$C'(R)/C(R)$ where $R$ is large and try to find an expression for this
ratio in terms of other known quantities. This is what we do in
the next section.

\subsection{Conservation laws}
\label{sec:conservation-laws}

We now show that energy conservation $\nabla_{\mu}T^{\mu 0}=0$ and conservation of the soft current
\[
j^{\mu} = F^{\mu\nu}\del_{\nu}\lambda
\] 
 is enough to fix the leading low energy absorption rate $\Gamma_{\rm abs}$ for electromagnetic radiation uniquely. We use the same contour as depicted in Fig.~\ref{fig:1}. From the conservation of the soft current it follows that
\[\label{eq:softcons}
0 \wkeq \oint_{C} \star j = \int_{r_H}^R g^{tt}g^{ij}F_{ti}\del_{j}\lambda\Big|^T_{-T}\sqrt{-g} d\Omega dr 
+ \int_{-T}^T g^{rr}g^{AB}F_{rA}\del_{B}\lambda\sqrt{-g} d\Omega dt\Bigg|^R_{r_{H}}.
\] 
Using the field strength for the $+$ mode \eqref{eq:goodsolution} and 
\[
\lambda(r,\theta) = \sum_{\ell ,m} c_{\ell  m}\Lambda_{\ell  m}(r)Y_{\ell m}(\theta),
\] 
we find for the time slices and a particular mode of the photon depending on the parameters $(\omega,\ell,m)$ that
\[
\int g^{rr}g^{AB}F_{rA}\del_{B}\lambda\sqrt{-g} d\Omega dt = -2\ell (\ell +p-1)\omega \sin(\omega T) \frac{\sqrt{r^{2p}f(r)g(r)^{p-1}}\Lambda_{\ell m}(r)C'(r)}{\ell (\ell +p-1)f(r)-r^{2}\omega^{2}g(r)}\Big|_{r_{H}}^{R}.
\] 
This is to be evaluated for the two cases $r=r_{H}+\delta\to r_{H}$ and $r=R\to\infty$. No approximations have been made at this point. We also already performed the integration over the sphere by making use of the orthogonality relation for vector spherical harmonics on $S^{p}$
\[
\int d\Omega_{p} \sqrt{\gamma}\gamma^{AB}\nabla_{A}Y_{\ell m}(\theta^{A})\nabla_{B}Y_{\ell 'm'}^{*}(\theta^{A}) = -\ell (\ell +p-1)\delta_{\ell \ell '}\delta_{mm'}.
\] Note that the indices $m$ and $m'$ are multi-indices, the length of which depending on $p$.

We continue to examine the contour integral \eqref{eq:softcons}. The integrals over the spatial slices combine to become
\begin{align}
\int_{r_{H}}^{R}g^{tt}g^{ij}&F_{ti}\del_{j}\lambda\sqrt{-g} \Big|^{T}_{-T} d\Omega dr\nln \qquad\qquad&= 2\ell (\ell+p -1)\omega \sin \omega T \int_{r_{H}}^{R} \sqrt{r^{2p}f(r)g(r)^{p-1}} \left(\frac{C(r)\Lambda_{\ell m}(r)}{f(r)r^{2}}+\frac{C'(r)\Lambda'_{\ell m}(r)}{\ell (\ell +p-1)f(r)-r^{2}\omega^{2}g(r)}\right)
\end{align}
By integrating the derivative on $C(r)$ in the second term by parts, the right hand side can be turned into a total derivative and a bulk part which we interpret as a second order differential equation for $\Lambda_{\ell m}$. The bulk part is proportional to the $C$ equation of
motion~\eqref{eq:C-DE} operator applied to $\Lambda_{\ell m}(r)$. We would like
the bulk term to become subleading in $\omega$, which we can achieve
by demanding that $\Lambda$ solve the $\omega = 0$ equation of motion
for $C$. Thus, $\Lambda$ plays the role of the ``intermediate
solution'' in the matched asymptotic expansion approach, see
e.g.~\cite{Fabbri:1975sa, Fabbri:1977ux}. Then, after setting the bulk part to zero one finds
\begin{multline}
\int_{r_{H}}^{R}g^{tt}g^{ij}F_{ti}\del_{j}\lambda\sqrt{-g} \Big|^{T}_{-T} d\Omega dr = 
2\ell (\ell +p-1)\omega sin(\omega T)\frac{r^p\sqrt{f(r) g(r)^{p-1}}}{\ell (\ell +p-1)f(r) - \omega^2 r^2 g(r)} C(r)\Lambda'(r)\bigg|_{r_H}^R
\end{multline}
 Finally, we combine the boundary pieces from the spatial and temporal slices. Upon taking the small $\omega$ and small $\delta$ limit according to the previously defined prescription, the conservation law constrains the form of the solution via
\begin{equation} \label{eq:sconsres}
\frac{\omega^{2}R^{p}}{\ell(\ell+p-1)}\begin{vmatrix} C(R) & C'(R) \\ \Lambda(R) & \Lambda'(R)\end{vmatrix} = -\sqrt{\frac{f(r_{H})}{g(r_{H})}} R_{H}^{p-2}\begin{vmatrix} C(r_{H}) & C'(r_{H}) \\ \Lambda(r_{H}) & \Lambda'(r_{H})\end{vmatrix} + \mathcal{O}(\omega^{3}).
\end{equation} where we recognize \[W(C,\Lambda) = \big(C(r)\Lambda'(r) - C'(r)\Lambda(r)\big) = \begin{vmatrix} C(r) & C'(r) \\ \Lambda(r) & \Lambda'(r)\end{vmatrix}\]
as the Wronskian determinant. The appearance of the Wronskian between $C(r)$ and $\Lambda(r)$, with $\Lambda$ a solution of~\eqref{eq:C-DE} can also be derived directly from
the differential equation~\eqref{eq:C-DE} as shown in~\cite{Castro:2013lba}, see appx~\ref{sec:wronskian}. We have shown
that the conservation of the Wronskian follows from large gauge symmetry. 

After deriving \eqref{eq:sconsres}, we also need to find a relation from conservation of energy. For this, we need two components of the energy-momentum tensor
\[\oint_{C}\star J =  \left(\int_{\Sigma_{-}} - \int_{\Sigma^{+}}\right)g^{tt}T_{tt}\Big|_{t=\pm T}\sqrt{-g} d\Omega dr + \left(\int_{H} - \int_{\Lambda}\right) g^{rr}T_{rt}\sqrt{-g}\Big|_{t=r_{H},R} d\Omega dt\label{eq:conenergphot}\]
which are given by
\begin{align}
  \label{eq:1}
  T_{rt} &= \phantom{-}\int \frac{d\omega_{1}d\omega_{2}}{4\pi^{2}} g^{AB}A_{A}^{\omega_{1}}F_{rB}^{\omega_{2}} (i\omega_{1})e^{-i(\omega_{1}+\omega_{2})t}\\
  T_{tt} &= \frac12 \int \frac{d\omega_{1}d\omega_{2}}{4\pi^{2}} \Big(A_{i}^{\omega_{1}}A_{j}^{\omega_{2}}g^{ij} \omega_{1}\omega_{2} - f(r) g^{ij}g^{kl} F_{ik}^{\omega_{1}}F_{jl}^{\omega_{2}}\Big) e^{-i(\omega_{1}+\omega_{2})t}
\end{align}
Again, the integration over spatial slices $\Sigma_{\pm}$ conspires to produce a factor of $(\omega_{1} + \omega_{2}) \frac{\sin (\omega_{1}+\omega_{2})T}{(\omega_{1}+\omega_{2})\pi}$ while the integrations over the time slices gives only a factor of $\frac{\sin (\omega_{1}+\omega_{2})T}{(\omega_{1}+\omega_{2})\pi}$. Thus the spatial slices do not actually contribute to the calculation. It follows that
\[\int d\Omega g^{rr} T_{tr}\sqrt{-g}\Big|_{r=r_{H}} = \int d\Omega g^{rr} T_{tr}\sqrt{-g}\Big|_{r=R}\] or, more specifically 
\begin{align}
\int d\Omega \sqrt{\gamma} \sqrt{\frac{f(r_{H})}{g(r_{H})}R_{H}^{2(p-2)}} &\gamma^{AB} F_{tA}^{\omega_{1}}(r_{H},\Omega) F_{rB}^{-\omega_{1}}(r_{H},\Omega)\nln \qquad\qquad&=\int d\Omega \ \sqrt{\gamma}\ R^{p-2} \gamma^{AB} F_{tA}^{\omega_{1}}(R,\Omega) F_{rB}^{-\omega_{1}}(R,\Omega).
\end{align}
We find an expression which holds for all $\omega$
\[
R^{p}\frac{C(R)C'(R)}{\ell (\ell +p-1)-\omega^{2}R^{2}} = \sqrt{\frac{f(r_{H})}{g(r_{H})}}R_H^p\frac{C(r_{H})C'(r_{H})}{\ell (\ell +p-1)f(r_{H})-\omega^{2}R_{H}^{2}}.\label{eq:energyconserv}
\] 
Use of the definition $R_{H}^{2} =
r_{H}^{2}g(r_{H})$ made the above relations slightly more aesthetically pleasing. Upon taking the small $\omega$ limit as well as the near-horizon limit -- where we strictly adhere to our prescription $\delta\ll \omega r_{H}^{2}$ to avoid order of limits issues --  we drop the $f$
term in the denominator so that the two conservation laws take the form
\begin{subequations}\label{eq:cons-system}\begin{align}
\frac{\omega^{2}R^{p}}{\ell(\ell+p-1)}\begin{vmatrix} C(R) & C'(R) \\ \Lambda(R) & \Lambda'(R)\end{vmatrix} = -\sqrt{\frac{f(r_{H})}{g(r_{H})}} R_{H}^{p-2}\begin{vmatrix} C(r_{H}) & C'(r_{H}) \\ \Lambda(r_{H}) & \Lambda'(r_{H})\end{vmatrix}\\
\frac{\omega^{2}R^{p}}{\ell(\ell+p-1)}C(R)C'(R) = -\sqrt{\frac{f(r_{H})}{g(r_{H})}}C(r_{H})C'(r_{H})R_{H}^{p-2}
\end{align}\end{subequations}

Now recall that $\Lambda_{\ell m}(r)$ should solve the $\omega = 0$ equation of
motion for $C$, elsewhere also called the intermediate solution. Since this is a second order differential equation, there are two solutions. Let us choose the solution
that falls off at large $r$ like $r^{-(\ell+1)}$ and note that
$\Lambda_{\ell m}(r_H)$ is finite and $\Lambda_{\ell m}'(r_H) = 0$. In the limit
$\delta \ll \omega r_H^2, \omega r_H\ll \omega R \ll 1$, we find a quadratic equation for $C'(R)/C(R)$ from~\eqref{eq:cons-system}:
\begin{equation}
\frac{C(R)}{C'(R)}\left(\frac{\ell+p-1}{R} + \frac{C'(R)}{C(R)}\right)^2
  = \frac{i\ell (\ell +p-1)R_{H}^{p-2}}{\omega R^P} \frac{\Lambda(r_H)^2}{\Lambda(R)^2}.
\end{equation}
We already inserted \eqref{eq:nearratio} into this equation. In the limit we are considering the right hand side should be treated
as large, in which case there are two solutions. The non-spurious solution is
\begin{equation}
\frac{C'(R)}{C(R)} \approx -\frac{\ell +p-1}{R} + i\frac{\ell (\ell +p-1)R_{H}^{p-2}}{\omega R^p} \frac{\Lambda(r_H)^2}{\Lambda(R)^2}.
\end{equation}
This is the relation we promised to derive in
sec.~\ref{sec:grad-mode-sols}. We can now plug this last result
into~\eqref{eq:C2divC1p} to get
\begin{equation}\label{eq:endratio}
\frac{C_{2}}{C_{1}} = -\left(\frac{\omega R}{2}\right)^{2\ell+p-1}\frac{\pi (\ell +p -1)(2\ell+p-1)}{2\ell \Gamma\left(\ell + \frac{p+1}{2}\right)^{2}}\left(1 + i\frac{\omega R^{p-1}}{R_{H}^{p-2}}\frac{2\ell+p-1}{\ell(\ell+p-1)}\right)\frac{\Lambda(r_{H})^{2}}{\Lambda(R)^{2}}
\end{equation}
With this, we are now in a position to examine the absorption rate. The ratio $\frac{C_2}{C_1}$ is very small, thus taking \eqref{eq:abs-21}, \eqref{eq:endratio}, and \eqref{eq:zeroratio} yields the absorption rate
\[\label{eq:final-rate}
\Gamma_{\rm abs} = \left(\frac{\omega R_{H}}{2}\right)^{2\ell+p}\frac{\pi^{2}(2\ell+p-1)^{2}}{\ell^{2}\Gamma\left(\ell + \frac{p+1}{2}\right)^{2}}\frac{\Gamma\left(\frac{\ell}{p-1} + 2\right)^{2}}{\Gamma\left(\frac{\ell}{p-1} + \frac{3}{2}\right)^{2}}g(r_{H})^{-\ell}
\]

The final solution is correct even for Reissner--Nordstr\"om-type
black holes in higher dimensions. The factor $g(r_{H})$ takes care of
the parameter dependence. The function $g(r)$ for $p+2$ dimensional RN
black holes is given in the appendix. In fact, this is should be the
general result for any spherically symmetric black hole solution. The
form given here faithfully reproduces the result for the Schwarzschild
solution in four dimensions when using that $g(r_{H}) = 16$. One can
recover the correct scaling for the extremal limit, in the usual
way eg.~\cite{Gubser:1997cm}, by tuning $\omega$ such that
$r_H^{p-1}\sim \omega R_H$ and thus
$g(r_H)\sim (\omega R_H)^{-\frac{2}{p-1}}$, which agrees with results
in~\cite{Crispino:2000jx, Gubser:1997cm}. If one does not take the
scaling limit the absorption probability goes to zero in the extremal
case, as in the case for minimally coupled fermions~\cite{Das:1996we}.

\section{Discussion}

We have shown that the leading low energy photon absorption rate of
black holes is fixed by large gauge invariance and conservation of
energy. At this point let us comment on the relationship of our
calculation to other approaches.

First, let us compare with doing the calculation via matched
asymptotic expansions. One can do the calculation in this way, and it
involves solving the three limits used above. However, the
interpretation and method is conceptually different. We have
conservation laws that identify constants of motion between two fixed
radii. The $\omega = 0$ solution gives a parameter ($\Lambda$) for the
gauge conservation law. In a matched asymptotic analysis, one instead
expands in the three regions and then matches the coefficients of
asymptotic behavior between large $r$ in the near-horizon region and
small $r$ in the intermediate region, and large $r$ in the
intermediate region and small $r$ in the flat region. Each matching
condition is \emph{two} conditions on the solution, whereas each
conservation law is only \emph{one} condition.

Second, let us comment on the connection to
HPS~\cite{Hawking:2016msc}. At first glance, our calculation looks
quite different from the discussion of HPS: we have regulated the
calculation on spatial and timelike surfaces instead of working near
null infinity; moreover, our order of limits is such that we never
reach null infinity; and finally, our calculation is entirely at the
level of the classical equations of motion. In fact, these are all
superficial distinctions. Our result can be interpreted in the
following way. We solve the classical equations of motion for large r
and the near-horizon region. This allows us to canonically quantize
using asymptotically flat modes and near-horizon modes. Then, we argue
that there are two conservation laws that fix the Bogolyubov
transformation between these two sets of modes. This is nothing but a
Ward identity, since one could now use the Bogolyubov transformation
to evaluate correlators with interior insertions. To wit, we have just
worked out in explicit detail the HPS Ward identity in the background
of a non-evaporating black hole.

There is one more point that deserves further exposition: the role of
the $\omega = 0$ equation. We found that in order to have a useful
conservation law, i.e., such that we can drop the spacelike parts of
the contour, we needed the gauge parameter in the conservation law to
satisfy the $\omega = 0$ photon equation of motion. We motivated this
result by observing that it is only in that case that the
longitudinal mode degenerates with the $\omega \to 0$ limit of the
physical mode. But, one might ask, why didn't this complication arise
when deriving Weinberg's soft theorem from large gauge
transformations? In fact, our conservation law needs only the
relationship of the gauge parameter between the two boundary surfaces,
$r=R$ and $r=r_H$. For the soft theorem, this is replaced by the
relationship between the gauge parameter on $\scri^-$ and $\scri^+$ for
 a longitudinal mode that behaves like the $\omega \to 0$
limit of the transverse mode. That is to say, the $\omega =0$ equation
is the analogue of the antipodal identification in~\cite{Strominger:2013jfa,Strominger:2013lka}, for this
calculation.

Relatedly, one may wonder about the universality of the photon
calculation, since we solved a differential equation that depended on
the geometry away from the horizon and the flat limits. In fact, we
expect the suggestive form given in~\eqref{eq:final-rate} is
universal. First, since this should be directly related (via small
gauge transformation and spherical harmonic decomposition) to the
antipodal identification used in~\cite{Hawking:2016msc}. More
explicitly, when $\omega = 0$, unlike for $\omega \neq 0$, the equation
has three regular singular points at $r=0$, $r=r_H$ and $r=\infty$;
hence the ${}_2F_1$ solution. The solution is completely determined by
the behavior at these points. One might worry that $r=0$ could be a
source of nonuniversality; however, in these coordinates the
continuation to $r < r_H$ does not describe the black hole interior but
rather a second copy of the asymptotic flat region. That is $r=0$ is a 
second copy of $r=\infty$. One can see this explicitly
from~\eqref{eq:sch-coord-trans} and~\eqref{eq:rn-coord-trans} in the
appendix, which have a $r\mapsto \frac{r_H^2}{r}$ inversion
symmetry. Thus, the equation is entirely determined by the near-horizon and
asymptotically flat physics.

If one develops our approach in AdS, then we expect that one may
interpret the $\lambda$ equation as a flow for a membrane paradigm, in
the spirit of~\cite{Iqbal:2008by}. We leave that intuition for future
investigations. Relatedly, it might be interesting to revisit the
effective string calculations for absorption (and emission)
rates~\cite{Strominger:1996sh, Callan:1996dv, Dhar:1996vu, Das:1996wn,
  Das:1996jy, Maldacena:1996ix, Gubser:1997cm}, perhaps using more modern AdS/CFT
technology from~\cite{Avery:2009tu}.

Finally, we would like to emphasize that our approach did not depend on the spherical symmetry of the problem. The advantage of investigating this particular set of black hole solutions is that the equations of motion separate. However, absorption rates have been studied and are known explicitly for the uncharged Kerr black hole \cite{Page:1977um}, though not for the Kerr--Newman solution to the authors' knowledge. Additionally, we concentrated on the very simple case of the minimally coupled photon. A natural expansion of this work is to investigate absorption rates for gravitational waves and their possible relation to Strominger's ${\rm BMS}^0$ symmetry \cite{Strominger:2013jfa}. We will leave these problems for future work.

\paragraph{Acknowledgments}

SGA would like to thank OSU, CERN, and Harvard for their hospitality
while working on this paper. SGA is grateful for conversations with
Samir Mathur at this project's inception. SGA benefitted from
discussions with Borun Chowdhury and Miguel Paulos, the latter also
giving feedback on an early draft. SGA was supported in part by the
Office of the Vice-President for Research and Graduate Studies at MSU,
and was supported by US DOE grant de-sc0010010 at Brown
University. BUWS was supported by the Center for Mathematical Sciences
and Applications at Harvard University, the Cheng Yu-Tung fund, and
NSF grant 1205550. BUWS is thankful to Andrew Strominger, Alexander
Zhiboedov, and Thomas Dumitrescu for discussions.

\appendix

\section{Schwarzschild and RN Metrics in DGM Coordinates}
\label{sec:schw-rn-metr}

The Schwarzschild metric in $p+2$ dimensions is given by 
\[
ds^{2} = -f(r')dt^{2}+f(r')^{-1}dr^{'2} + r^{'2}d\Omega_{p}
\] 
where $f(r') = 1 - \left(\frac{r_{s}}{r'}\right)^{p-1}$. A coordinate transform to the form \eqref{eq:metric} is 
\[\label{eq:sch-coord-trans}
r'(r) = \frac{(r^{p-1} + r_H^{p-1})^\frac{2}{p-1}}{r}
\] 
and the two functions $f(r)$ and $g(r)$ are found to be
\[
f(r) = \left(\frac{r^{p-1} - r_{H}^{p-1}}{r^{p-1}+r^{p-1}_{H}}\right)^{2},\qquad\qquad 
g(r) = \frac{(r^{p-1} + r_H^{p-1})^\frac{4}{p-1}}{r^4},\qquad\qquad 
r_{H}^{p-1} = \frac{r_{S}^{p-1}}{4}.
\]  
An integration constant has been chosen such that the two functions satisfy the asymptotically flat condition 
\[
\lim_{r\to\infty}f(r)=\lim_{r\to\infty}g(r)=1.
\] 
Note that $g(r)$ is finite at the horizon $g(r_{H}) = 2^{\frac{4}{p-1}}$ and in particular 
\[
R_{H}^{2} = r_{H}^{2}g(r_{H}) = r_{S}^{2}.
\]
Similarly, for the Reissner-Nordstr\"om metric with function 
\[
f(r') = 1-\left(\frac{r_{S}}{r'}\right)^{p-1} + \left(\frac{r_{Q}}{r'}\right)^{2p-2}
\] 
we can give a coordinate transform 
\[\label{eq:rn-coord-trans}
r'(r) = \left(\frac{(4 r^{p-1}+r_{S}^{p-1})^{2}-4 r_{Q}^{2p-2}}{16 r^{p-1}}\right)^{\frac{1}{p-1}}
\]
The metric is still given by the general form above \eqref{eq:metric}, but with the two functions 
\begin{align}\label{eq:general-rn-fg}
f(r) &= \frac{[4 (4r^{2(p-1)} + r_{Q}^{2(p-1)}) -r_{S}^{2(p-1)}]^{2}}{(4 r^{p-1}-2r^{p-1}_{Q}+r^{p-1}_{S})^{2}(4 r^{p-1}+2r^{p-1}_{Q}+r^{p-1}_{S})^{2}}\\
g(r) &= \left(\frac{(4r^{p-1}-2r^{p-1}_{Q}+r^{p-1}_{S})(4r^{p-1}+2r^{p-1}_{Q}+r^{p-1}_{S})}{16 r^{2(p-1)}}\right)^{\frac{2}{p-1}}.
\end{align} 
Here $r_{Q}^{2} = \frac{Q^{2}}{4\pi}$ is the charge of the black hole. In these coordinates, the horizons are located at \[r_{H}^{p-1} = \pm\frac14\sqrt{r_{S}^{2p-2}-4r_{Q}^{2p-2}}\] such that $f(r)\propto (r^{2}-r_{H}^{2})^{2}$ and $g(r_{H})$ is, again, finite. The extremal Reissner--Nordstr\"om black hole is obtained in the limit $r_{S} = 2 r_{Q}$. In our coordinates, $g(r)$ diverges at the horizon.

\section{Solenoidal Mode}
\label{sec:solenoidal-mode}

For completeness, the solenoidal mode equation of motion takes the form
\begin{equation}
D''(r) + \frac{1}{2}\left((p-3)\frac{g'(r)}{g(r)} + 2\frac{p-2}{r} + \frac{f'(r)}{f(r)}\right)D'(r) + \left(\omega^2\frac{g(r)}{f(r)} - \frac{\ell(\ell + p -1) + (p-2)}{r^2}\right)D(r) = 0.
\end{equation}
The field strength is given by
\begin{equation}\begin{aligned}
F_{tr} &= 0\\
F_{tA} &= -i\omega D(r)\, \hat{W}^{(\ell)}_A e^{-i\omega t}\\
F_{rA} &= D'(r)\, \hat{W}^{(\ell)}_A e^{-i\omega t}\\
F_{AB} &= D(r)\, \big(\hat\nabla_{A}\hat{W}^{(\ell)}_B - \hat{\nabla}_B\hat{W}^{(\ell)}_A\big)
  e^{-i\omega t}
\end{aligned},\end{equation} Unlike the gradient mode, this
polarization does not have vanishing field strength as $\omega\to 0$,
so we do not expect it to mix with the residual gauge mode.

\section{Conservation of the Wronskian}
\label{sec:wronskian}

This appendix demonstrates how the conservation of the Wronskian
follows from the differential equation, following the discussion and
notation in~\cite{Castro:2013lba}. The radial equation of
motion~\eqref{eq:C-DE} may be written in the form
\begin{equation}
\frac{\drm}{\drm r}\left(\frac{r^p\sqrt{f(r) g(r)^{p-1}}}{\ell(\ell+p-1)f(r)-\omega^2 r^2g(r)}C'(r)\right) - r^{p-2}\sqrt{\frac{g(r)^{p-1}}{f(r)}}C(r) = 0.
\end{equation}
Following~\cite{Castro:2013lba}, define the two component vector
\begin{equation}
\Psi(r) = \begin{pmatrix} C(r)\\
\frac{r^p\sqrt{f(r) g(r)^{p-1}}}{\ell(\ell+p-1)f(r)-\omega^2 r^2g(r)}C'(r)
\end{pmatrix},
\end{equation}
which satisfies the first order differential equation
\begin{equation}
\Psi'(r) = A(r).\Psi(r)\qquad
A(r) = \begin{pmatrix} 
0 & \frac{\ell(\ell+p-1)f(r)-\omega^2 r^2g(r)}{r^p\sqrt{f(r) g(r)^{p-1}}}\\
r^{p-2}\sqrt{\frac{g(r)^{p-1}}{f(r)}} & 0
\end{pmatrix}.
\end{equation}
This two dimensional system has two linearly independent solutions;
call them $\Psi^{(1)}$ and $\Psi^{(2)}$. The \emph{fundamental matrix}
is the two-by-two matrix, $\Phi$, formed from
$(\Psi^{(1)}\, \Psi^{(2)})$. Formally one may solve the differential
equation for $\Phi(r)$ in the complex plane by writing a path-ordered
exponential
\[
\Phi(r) = \mathcal{P}\exp\left(\int_{r_0}^r A(z)\drm z\right) \Phi(r_0),
\]
and thus
\begin{equation}
\det\Phi(r) = \exp\left(\int_{r_0}^r \tr A(z)\drm z\right) \det\phi(r_0).
\end{equation}
From the vanishing trace of $A(r)$, $\tr A(r) = 0$, it follows that
$\det\Phi(r)$ is a constant of motion; and the determinant of the
fundamental matrix is nothing but (a prefactor times) the Wronskian. Forming the vectors
$\Psi^{(1)}$ and $\Psi^{(2)}$ from $C(r)$ and $\Lambda(r)$,
respectively, it follows that
\begin{equation}
\det \Phi = \frac{r^p\sqrt{f(r) g(r)^{p-1}}}{\ell(\ell+p-1)f(r)-\omega^2 r^2g(r)}\left[C(r) \Lambda'(r) - C'(r) \Lambda(r)\right]
\end{equation}
is a constant of motion. Applying our order of limits, one arrives
at~\eqref{eq:sconsres} in the main text. Let us emphasize that all
conservation laws, by their very nature, can be derived from the
equations of motion; the advantage of having a symmetry principle is
that one may say something without even looking at the detailed form
of the equations of motion.

\printbibliography
\end{document}